\documentclass[onecolumn,nofootinbib]{revtex4}
\usepackage{graphicx}
\usepackage{amsmath}
\usepackage{amssymb}
\usepackage{float}
\usepackage[all]{xy}
\usepackage{amsfonts}
\usepackage{dcolumn}
\usepackage{bm}
\usepackage{xcolor}
\usepackage{caption}
\usepackage{hyperref}
\usepackage{multirow}
\usepackage[utf8x]{inputenc}
\usepackage{epstopdf}
\usepackage{capt-of}
\usepackage{dcolumn}   
\usepackage{bm}
\usepackage{dsfont}
\usepackage{relsize}
\usepackage{mathtools}

\begin{document}

\title{Baryogenesis in $f(P)$ Gravity}

\author{Snehasish Bhattacharjee \footnote{Email: snehasish.bhattacharjee.666@gmail.com}
  }\

\affiliation{Department of Physics, Indian Institute of Technology, Hyderabad 502285, India}
\date{\today}

\begin{abstract}
In this work, we investigate gravitational baryogenesis in the framework of $f(P)$ gravity to understand the applicability of this class of modified gravity in addressing the baryon asymmetry of the Universe. For the analysis, we set $f(P) = \alpha P$ where $\alpha$ is the model parameter. We found that in $f(P)$ gravity, the CP-violating interaction acquires a modification through the addition of the nontopological cubic term $P$ in addition to the Ricci scalar $R$ and the mathematical expression of the baryon-to-entropy ratio depends not only on the time derivative of $R$ but also the time derivative of $P$. Additionally, we also investigate the consequences of a more complete and generalized CP-violating interaction proportional to $f(P)$ instead of $P$ in addressing the baryon asymmetry of the Universe. For this type of interaction, we report that the baryon-to-entropy ratio is proportional to $\dot{R}$, $\dot{P}$ and $f^{'}(P)$.  We report that for both of these cases, rational values of $\alpha$ generate acceptable baryon-to-entropy ratios compatible with observations.

\end{abstract}

\maketitle
  
\section{Introduction}

Cosmological observations assert that the Universe is curvatureless and is currently experiencing a phase of accelerated expansion \cite{in1to6}. This phenomenon can be elegantly explained if two unknown entities are present in the Universe in abundance. Firstly, there is the dark energy that manifests at cosmological scales and is responsible for accelerating the Universe, and then there is the dark matter which manifests at smaller galactic and extragalactic scales and is responsible for the formation of the large scale structures. Nonetheless, no experimental evidence has been reported to directly confirm the existence of these entities. In this spirit, several alternating scenarios are being actively considered to explain these phenomena by modifying the theory of gravity.\\  
Modified theories of gravity have been successful in explaining numerous cosmological enigmas such as the flat galactic rotation curves and the cosmic acceleration without employing dark matter and dark energy \cite{in7,in8to10}. Although modified theories of gravity preserve the results and conclusions of GR at the Solar system scales, the gravitational interactions differ substantially at different scales \cite{in7}. GR, therefore, acts as a specific case of a more diverse and extended class of gravitational theories. \\   
In the last few years, there has been an expeditious development of a particular class of modified theory of gravity called the Einsteinian cubic gravity (ECG) where the gravitational interactions are governed through the cubic contractions of the Riemann tensor \cite{p19}. In \cite{p}, the authors generalized the ECG by substituting the Ricci scalar in the Einstein-Hilbert action with a suitable function of the curvature invariant $P$ which denotes the cubic contractions of the Riemann tensor. Thereafter, $f(P)$ gravity have been successfully applied to explain the late-time acceleration \cite{p,p12}, spherically symmetric black hole solutions \cite{p20to22}, and inflation \cite{p23}. Recently, tight constraints on $f(P)$ gravity have been reported from the energy conditions \cite{bhattap}. \\
It is quite well known and concurrently mysterious to note that the objects and structures in the Universe are composed of matter and that the Universe only contains a trifling amount of antimatter \cite{pdu1}. This phenomenon contrasts sharply with the predictions of the standard big bang cosmological model, according to which the Universe must have produced equal amounts of matter and antimatter only to be annihilated moments later. It is now believed that a minor but significant asymmetry surfaced between matter and antimatter which converted a small fraction of antimatter into the normal matter before they could annihilate. Thus, a tiny residual of matter remained and formed all the structures in the Universe. This phenomenon is also called baryon asymmetry or baryogenesis. Several promising theoretical frameworks have been reported to explain this enigmatic phenomenon by proposing interactions that are exotic and are also not restricted by the standard model \cite{pdu2}. \\
Gravitational baryogenesis is one of the most popular theoretical frameworks to explain baryogenesis in great detail \cite{pdu3}. In this framework, by employing the famous Sakharov conditions \cite{pdu4}, a CP-violating interaction was proposed which gives rise to a baryon asymmetry and can be represented as \cite{pdu3}
\begin{equation}\label{eq1}
\frac{1}{M_{*}} \int \sqrt{-g} \partial_{i}R J^{i} d^{4}x,
\end{equation}       
where $M_{*}$ represents the mass of the underlying effective theory, $R$ denotes the Ricci scalar, $g$ the metric determinant, and $J^{i}$ represents the baryonic current. Another interesting feature of this framework transpires from the fact that by replacing the Ricci scalar $R$ appearing in the CP-violating interaction (\ref{eq1}) with other curvature invariants, a finite and non-zero baryon asymmetry can be produced in the radiation dominated Universe \cite{pdu} and owing to which, Gravitational baryogenesis have been exhaustively studied in numerous modified gravity theories \cite{pdu5to9}.\\
In this work, we shall try to investigate the applicability of $f(P)$ gravity in addressing the baryon asymmetry of the Universe in the framework of Gravitational baryogenesis. \\
The manuscript is organized as follows: In Section \ref{sec2} we present an overview of $f(P)$ gravity. In Section \ref{sec3}, we investigate gravitational baryogenesis in $f(P)$ gravity and in Section \ref{sec4} we summarize the results and conclude the work.

\section{$f(P)$ Gravity}\label{sec2}

The nontopological cubic term $P$ is given by \cite{p}
\begin{multline}
P = \chi_{1} R^{3}+ \chi_{2} R^{c\rho}R^{d\sigma}R_{cd \rho \sigma}   + \chi_{3} R R^{cd} R_{cd}    +\chi_{4}R^{\rho \sigma}_{cd}R^{\gamma \delta}_{\rho \sigma} R^{cd}_{\gamma \delta}\\+  \chi_{5} R R_{cd\rho \sigma}R^{cd\rho \sigma}     +\chi_{6} R ^{\sigma \gamma}R_{cd\rho \sigma}R^{cd\rho \sigma}{}_{\gamma} +  \chi_{7} R^{d}_{c} R^{\rho}_{d}     R^{c}_{\rho}   + \chi_{8} R_{c}{}^{\rho}{}_{d}{}^{\sigma}R_{\rho}{}^{\gamma}{}_{\sigma}{}^{\delta}R_{\gamma}{}^{c}{}_{\delta}{}^{d} ,
\end{multline}

where $\chi_{i}$ are unknown parameters. To corroborate that for $f(P)$ gravity, the spectrum remain akin to that of the GR, the unknown parameters $\chi_{1}$, and $\chi_{3}$ are required to assume the following forms \cite{p19}

\begin{equation}
\chi_{1} = \frac{-1}{12}\left[ 5 \chi_{2}+ 24 \chi_{4} +  48 \chi_{5} +16 \chi_{6}+9\chi_{7}-3 \chi_{8}  \right], 
\end{equation}

\begin{equation}
\chi_{3} = \frac{1}{72}\left[3 \chi_{2} +36\chi_{4} + 64 \chi_{5}+ 22 \chi _{6}+ 9  \chi_{7}-6 \chi_{8}\right]. 
\end{equation}
       
Now, the action in $f(P)$ gravity is defined as \cite{p}

\begin{equation}\label{action}
\mathcal{S} = \frac{1}{2 \kappa}\int \sqrt{-g}\left( R + f(P)\right) d^{4}x, 
\end{equation}
where $R$ represents the Ricci scalar, $g$ represents the metric determinant, $\kappa = 8 \pi G$ represents the gravitational constant, and $f(P)$ represents a function of $P$.

Varying (Eq. \ref{action}) generates the field equation in $f(P)$ gravity as 
\begin{equation}\label{field}
G_{cd} = \kappa (T_{cd} +\bar{H}_{cd} ),
\end{equation}
where $T_{cd}$ represents the stress-energy-momentum tensor and reads
\begin{equation}
T_{cd} =  \left(\frac{-2}{\sqrt{-g}} \right) \left[\frac{\delta(\mathcal{L}_{m}\sqrt{-g})}{\delta g^{cd}} \right], 
\end{equation}
and 
\begin{equation}
\bar{H}_{cd}  = g_{cd}f(P)\nabla^{\epsilon}\nabla^{\varepsilon} \bar{\Pi}_{\epsilon(cd)\varepsilon}+R^{\epsilon\varepsilon\iota}(_{c}\bar{\Pi}_{d})_{\iota\epsilon\varepsilon} .
\end{equation}
The tensor $\bar{\Pi}_{\epsilon\varepsilon cd}$ is represented as $\Pi_{\epsilon\varepsilon cd} f^{'}(P)$, where $f^{'}(P) = df/dP$. Additionally, $\Pi_{\epsilon\varepsilon cd}$ is represented as \cite{p} 
\begin{equation}
\Pi_{\epsilon\varepsilon cd} = 12 \left[ 2g_{\varepsilon} (_{c}{}R_{\nu})_{\sigma \epsilon \iota} R ^{\iota \sigma} + \frac{1}{2} R_{\epsilon \varepsilon}^{\iota \sigma} R_{cd \iota \sigma}- 2 R _{\epsilon} (_{\epsilon}R_{\varepsilon})_{d} +6 R_{\epsilon}{}^{\iota}(_{c}{}^{\sigma}R_{\nu})_{\sigma \varepsilon \iota}  - 4R_{\iota}(_{c}g_{d})(_{\epsilon}R_{\varepsilon})^{\iota} -2 g_{\epsilon}(_{c}R_{d})_{\sigma \varepsilon \iota} R^{\iota \sigma}  \right]. 
\end{equation}

Assuming an FRW spacetime, the modified Friedmann equations are given by \cite{p}

\begin{equation}
3H^{2} = \kappa (\rho_{m} + \rho_{f(P)}),
\end{equation} 
\begin{equation}
2 \dot{H} + 3H^{2} = -\kappa( p_{m} + p_{f(P)}),
\end{equation}
where, $\rho_{f(P)}$, and $p_{f(P)}$ are represented as \cite{p}
\begin{equation}\label{rho}
\rho_{f(P)} = 18\chi H^{4} (\dot{H}+ H^{2} - H \frac{\partial}{\partial t} ) f^{'}(P)- f(P),
\end{equation}
\begin{equation}
p_{f(P)} = 6  \chi H^{3} \left[  H\frac{\partial^{2}}{\partial t^{2}}-5H\dot{H} +2(H^{2} + 2\dot{H})\frac{\partial}{\partial t}  - 3 H^{3}  \right] f^{'}(P)+f(P).
\end{equation}

Furthermore, for a FRW metric, $P = 6  \chi H^{4} (2 H^{2} + 3 \dot{H})$ where the parameter $\chi$ is given by 
\begin{equation}
\chi = 4 \chi_{4}+ 8 \chi_{5} + 2 \chi_{7} - \chi_{8}.
\end{equation}

\section{Gravitational baryogenesis}\label{sec3}

It is convenient to express the baryon asymmetry through the baryon-to-entropy ratio as it has been tightly constrained through observations which reads \cite{pdu1}
\begin{equation}
\frac{\eta_{B}}{s}\simeq 9 \times 10^{-11}.
\end{equation}
A key particular in baryogenesis is the Sakharov conditions \cite{pdu4} which must be obeyed. These conditions communicate that for baryogenesis to transpire, processes violating the baryon number, and the joint charge-parity interactions must be present and the entire phenomena must occur outside of thermal equilibrium.  \\
Due to the expansion of the Universe, the temperature $T$ descends continuously and after it drops below a critical value ($T_{D}$), the processes accountable for baryogenesis comes to a halt and the residual baryon-to-entropy ratio can be understood within the framework of gravitational baryogenesis through the following equation \cite{pdu3}  
\begin{equation}
\frac{\eta_{B}}{s}\simeq - \left(\frac{15 g_{b}}{4 \pi^{2} g_{s}} \right) \left[\frac{\dot{R}}{M_{*}^{2} T_{D}} \right], 
\end{equation}
where $g_{s}$, and $g_{b}$ denote respectively the total degrees of freedom for massless particles and baryons. Furthermore, in the radiation dominated Universe, by using the principles of statistical mechanics, we can express the energy density as a function of temperature $T$ through the following equation 
\begin{equation}\label{rhot}
\rho (T) = \frac{\pi^{2}}{30} g_{*} T^{4}.
\end{equation}  
We shall now try to investigate the viability of a simple $f(P)$ gravity model of the form $f(P) = \alpha P$ where $\alpha$ is a constant in addressing the baryon asymmetry. Additionally, we shall employ a power-law type scale factor of the form $a(t) \sim t^{m}$, where $m>0$ is a constant. We note that for $\alpha=0$, the field equations reduce to that of the GR. \\

\subsection{Baryogenesis in $f(P)$ gravity}

In $f(P)$ gravity, the CP-violating interaction (Eq.\ref{eq1}) undergoes a modification through the addition of the nontopological cubic term $P$ in addition to the Ricci scalar $R$ and the resultant interaction can be expressed as 
\begin{equation}\label{case1}
\frac{1}{M_{*}} \int \sqrt{-g} \partial_{i}(R+P) J^{i} d^{4}x.
\end{equation} 
Therefore, the residual baryon-to-entropy ratio in $f(P)$ gravity reads
\begin{equation}\label{main1}
\frac{\eta_{B}}{s}\simeq - \left(\frac{15 g_{b}}{4 \pi^{2} g_{s}} \right) \left[\frac{ \dot{R} + \dot{P} }{M_{*}^{2} T_{D}} \right]. 
\end{equation}

\begin{figure}[H]
\centering
  \includegraphics[width=8.5 cm]{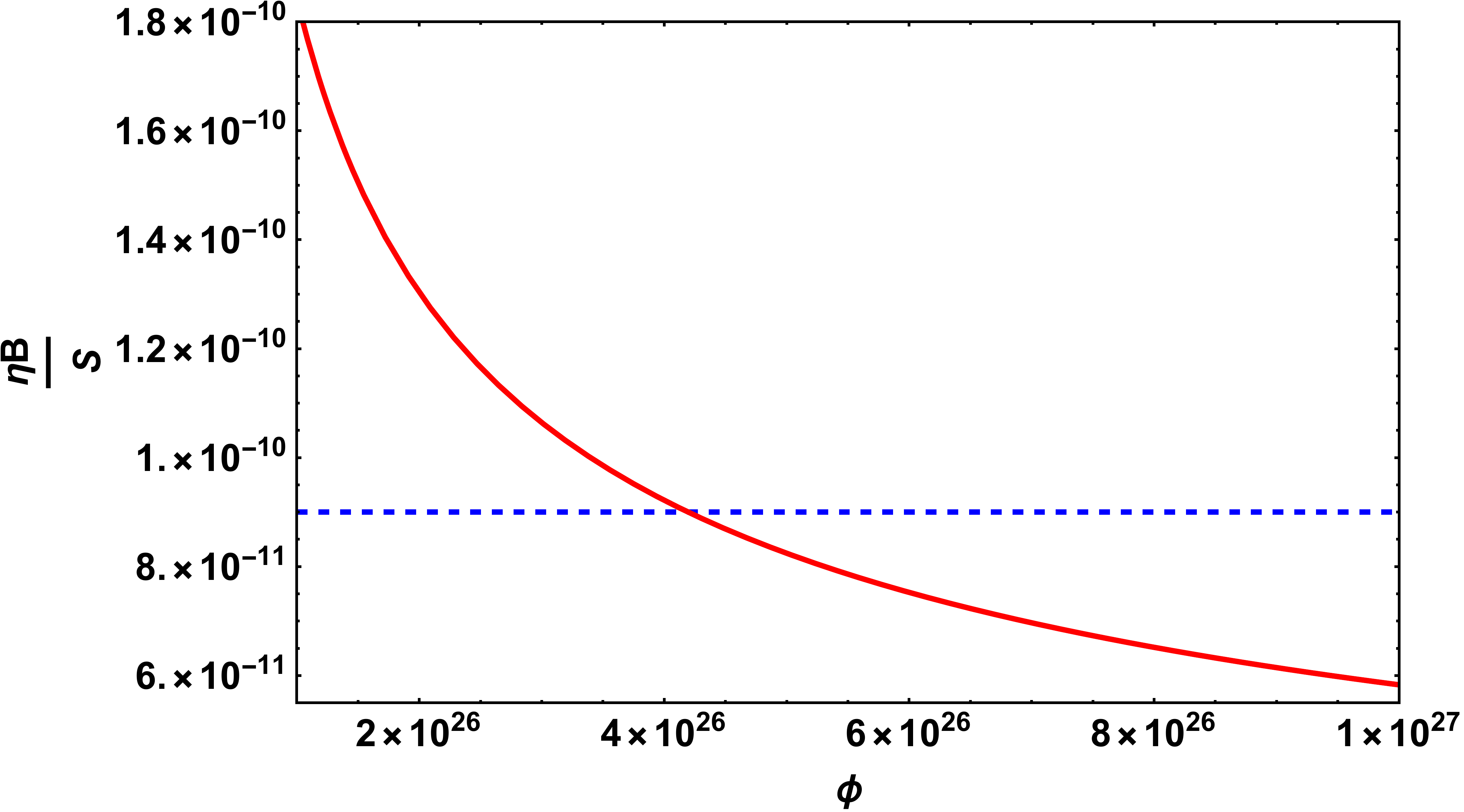} 
  \includegraphics[width=8.5 cm]{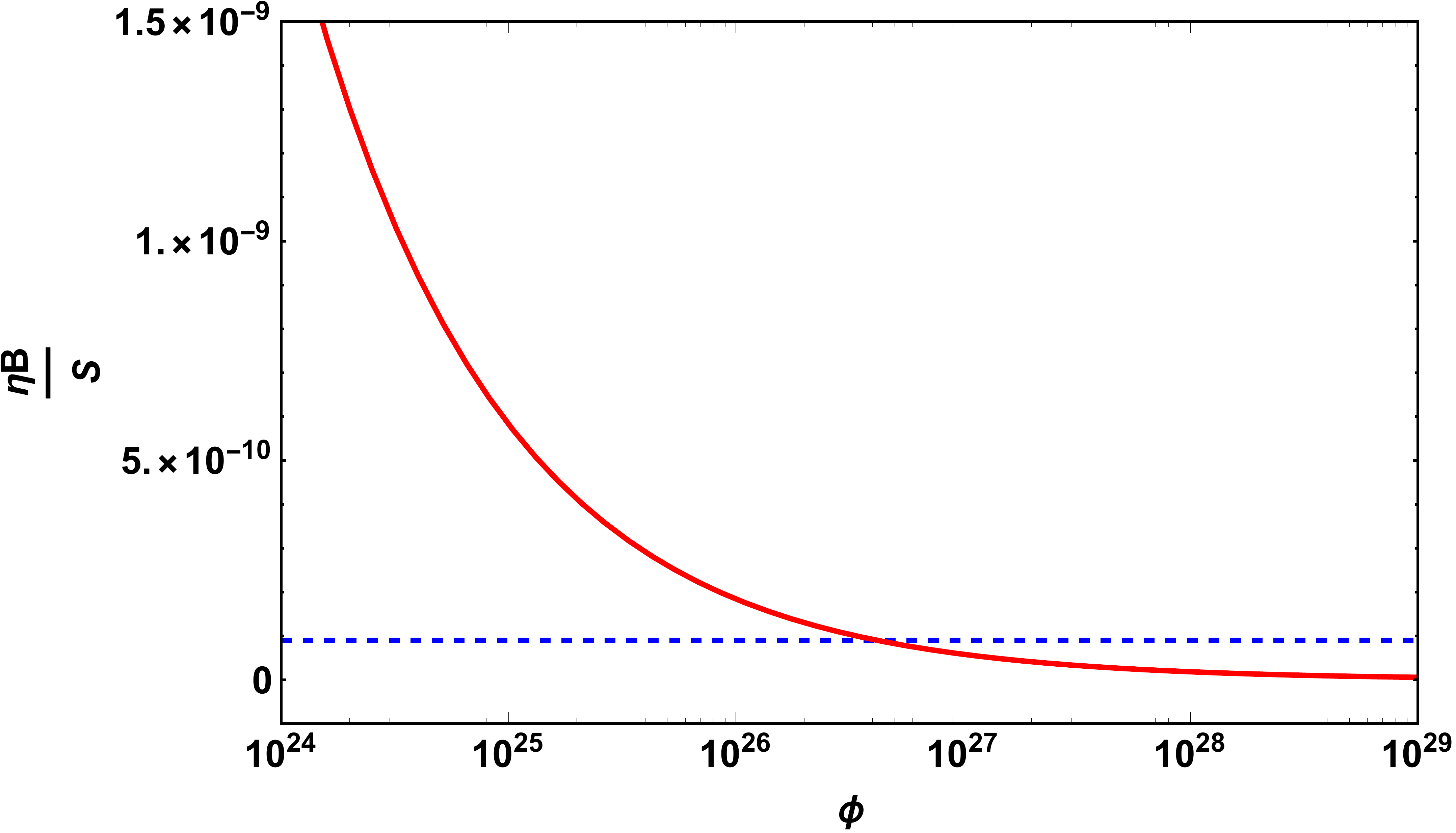} 
\caption{The left panel shows the profile of baryon-to-entropy ratio ($\frac{\eta_{B}}{s}$) as a function of $\phi$ in linear scale and the right panel shows the same in logarithmic scale. The figure is drawn for $m=1.5$, $T_{D}=2 \times 10^{12} GeV$, and $M_{*}=2\times 10^{16} GeV$.}
\label{fig1}
\end{figure}

Substituting $f(P) = \alpha P$ in Eq. \ref{rho}, the expression for the energy density ($\rho_{f(P)}$) reads 
\begin{equation}\label{rho.1}
\rho_{f(P)} = \frac{6 \alpha  \chi  m^6}{t^6}.
\end{equation} 
Let us denote the product $\alpha \chi$ by $\phi$. We shall proceed by equating Eq.\ref{rho.1} with Eq. \ref{rhot} to obtain an expression for the decoupling time ($t_{D}$) in terms of the critical temperature ($T_{D}$) as 
\begin{equation}\label{td}
t_{D}=\sqrt[6]{\frac{10}{53}} \sqrt[3]{\frac{3}{\pi }} \sqrt[6]{\frac{\phi  m^6}{T_{D}^4}}.
\end{equation}
Finally, we substitute Eq. \ref{td} in Eq.\ref{main1} to obtain the expression for the residual baryon-to-entropy ratio in the relevant $f(P)$ gravity model as
\begin{equation}
\frac{\eta_{B}}{s}\simeq \frac{m T_{D} \left(-30 \sqrt[3]{53} \left(\frac{\phi  m^6}{T_{D}^4}\right)^{2/3}+60 \sqrt[3]{53} m \left(\frac{\phi  m^6}{T_{D}^4}\right)^{2/3}+53\ 3^{2/3} \sqrt[3]{10} \pi ^{4/3} \phi m^4 (2 m-3)\right)}{4 \sqrt{10} 53^{5/6} \pi  M_{*}^2 \left(\frac{\phi  m^6}{T_{D}^4}\right)^{7/6}}.
\end{equation} 
Substituting $m=1.5$, $\phi=8 \times 10^{26}$, $T_{D}=2 \times 10^{12} GeV$, and $M_{*}=2\times 10^{16} GeV$, yields $\frac{\eta_{B}}{s}\simeq 5.9 \times 10^{-11}$ which equate nicely with observations. In Fig. \ref{fig1}, we show the profiles of $\frac{\eta_{B}}{s}$ as functions of $\phi$.
\subsection{Generalized baryogenesis interactions in $f(P)$ gravity}

\begin{figure}[H]
\centering
  \includegraphics[width=8.5 cm]{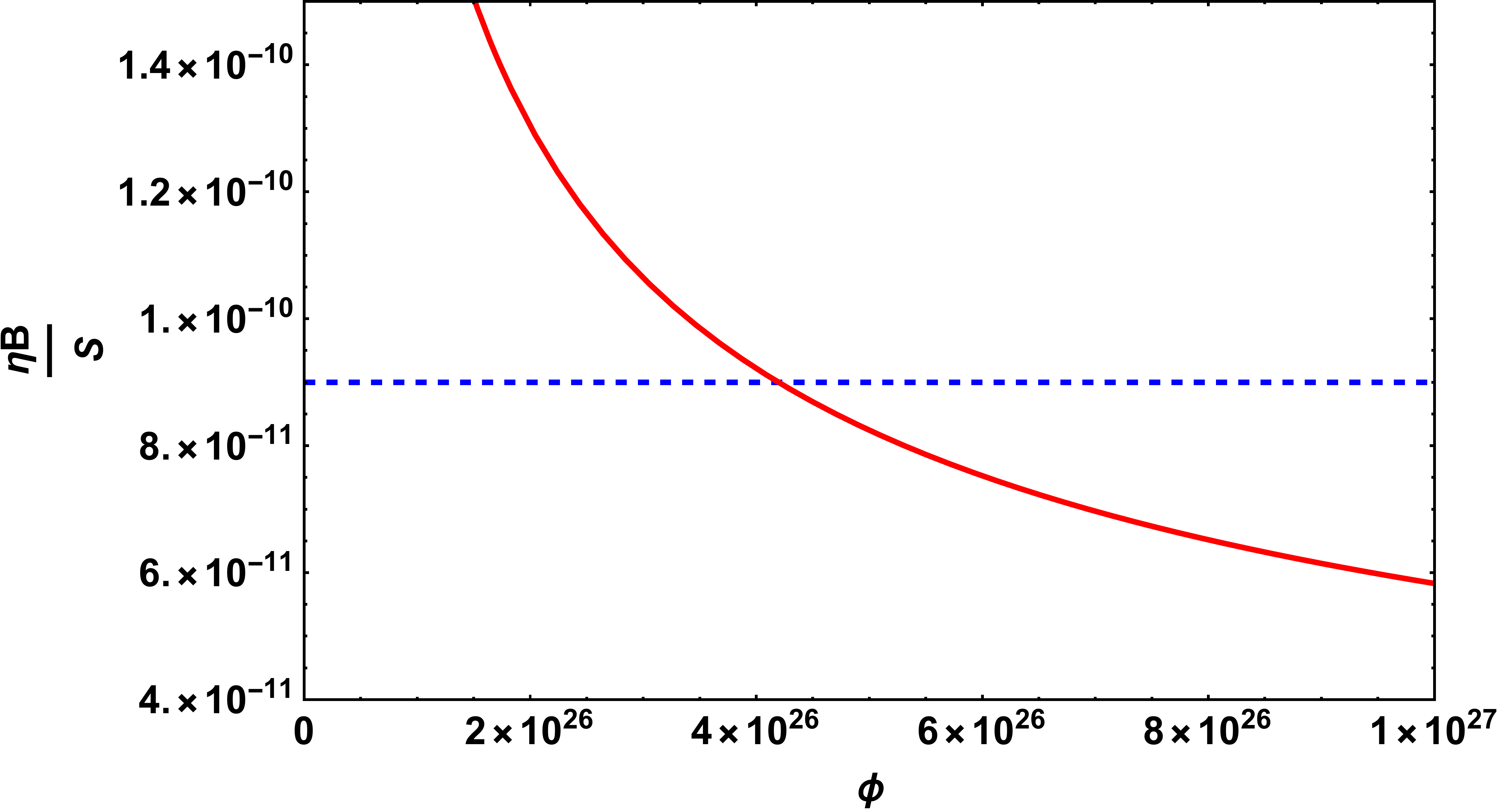}  
  \includegraphics[width=8.5 cm]{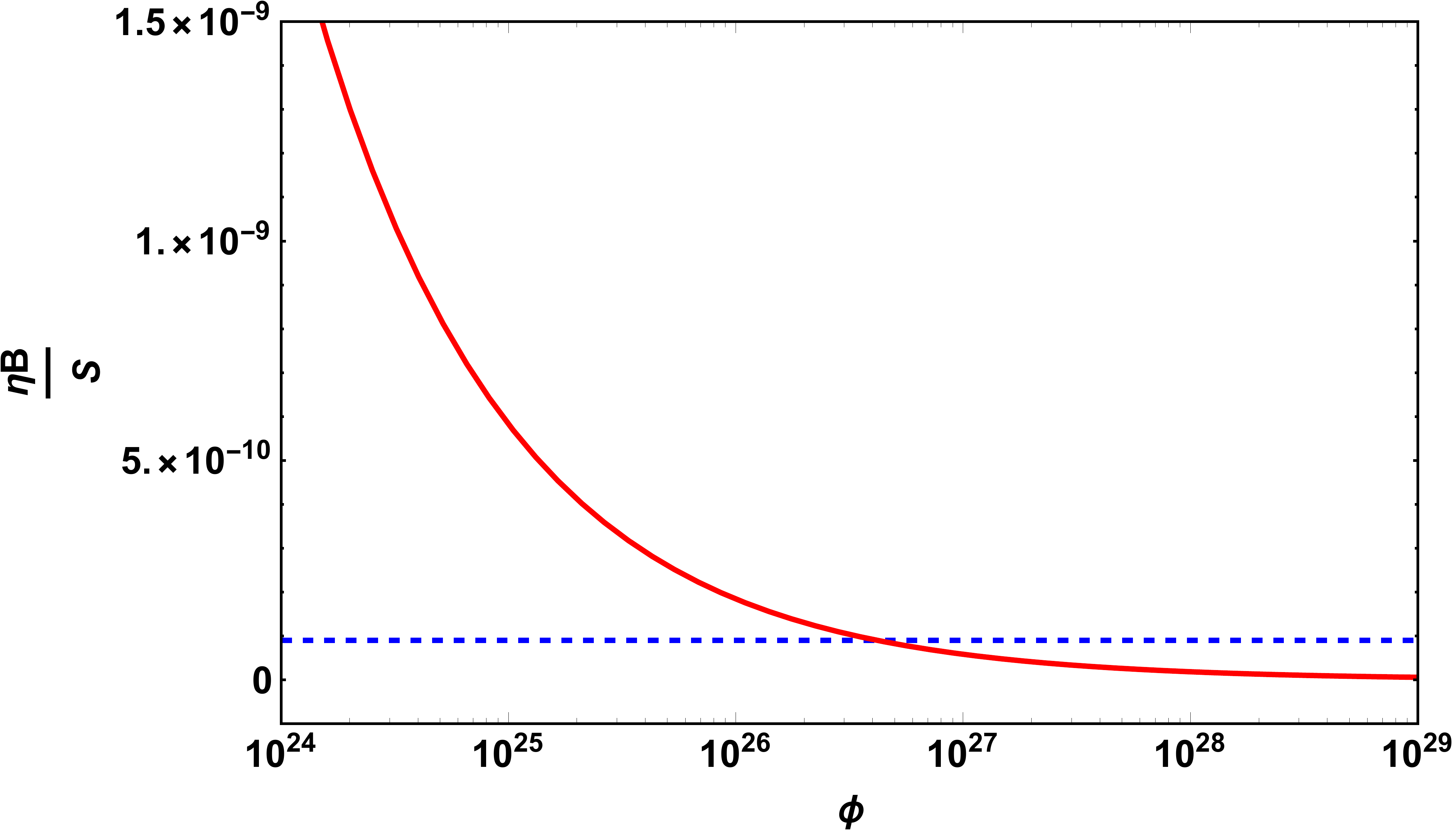} 
\caption{The left panel shows the profile of baryon-to-entropy ratio ($\frac{\eta_{B}}{s}$) as a function of $\phi$ in linear scale and the right panel shows the same in logarithmic scale for the generalized baryogenesis interaction (\ref{case2}). The figure is drawn for $m=1.5$, $T_{D}=2 \times 10^{12} GeV$, and $M_{*}=2\times 10^{16} GeV$.}
\label{fig2}
\end{figure}

We shall now try to investigate the consequences of a more complete and generalized CP-violating interaction proportional to $f(P)$ instead of $P$ in addressing the baryon asymmetry of the Universe for the chosen $f(P)$ gravity model. We can express the generalized CP-violating interaction in $f(P)$ gravity as 
\begin{equation}\label{case2}
\frac{1}{M_{*}} \int \sqrt{-g} \partial_{i}(R+f(P)) J^{i} d^{4}x.
\end{equation} 
The expression of baryon-to-entropy ratio therefore attains the following form 
\begin{equation}\label{main2}
\frac{\eta_{B}}{s}\simeq - \left(\frac{15 g_{b}}{4 \pi^{2} g_{s}} \right) \left[\frac{ \dot{R} + \dot{P}f^{'}(P) }{M_{*}^{2} T_{D}} \right]. 
\end{equation}
Likewise to the previous case, we shall now substitute Eq.\ref{td} in Eq.\ref{main2} to obtain the expression of baryon-to-entropy ratio for the model $f(P)=\alpha P$ as 
\begin{equation}
\frac{\eta_{B}}{s}\simeq\frac{m T_{D} \left(-30 \sqrt[3]{53} \left(\frac{\phi  m^6}{T_{D}^4}\right)^{2/3}+60 \sqrt[3]{53} m \left(\frac{\phi  m^6}{T_{D}^4}\right)^{2/3}+53\ 3^{2/3} \sqrt[3]{10} \pi ^{4/3} \phi  m^4 (2 m-3)\right)}{4 \sqrt{10} 53^{5/6} \pi  M_{*}^2 \left(\frac{\phi m^6}{T_{D}^4}\right)^{7/6}}.
\end{equation}
Substituting $m=1.5$, $\phi=5 \times 10^{26}$, $T_{D}=2 \times 10^{12} GeV$, and $M_{*}=2\times 10^{16} GeV$, gives $\frac{\eta_{B}}{s}\simeq 8.3 \times 10^{-11}$ which matches up well with the observational estimate. In Fig. \ref{fig2}, we show the profiles of $\frac{\eta_{B}}{s}$ as a function of $\phi$ for the generalized baryogenesis interaction (\ref{case2}).

\section{Conclusions}\label{sec4}

Einsteinian cubic gravity (ECG) is a modified theory of gravity where the gravitational interactions are governed through the cubic contractions of the Riemann tensor \cite{p19} instead of the usual Ricci scalar as in GR and has undergone substantial development in recent years. $f(P)$ gravity is a more complete version of ECG introduced in \cite{p} where the authors generalized the ECG by substituting the Ricci scalar in the Einstein-Hilbert action with a suitable function of the curvature invariant $P$ which denote the cubic contractions of the Riemann tensor. Since then, $f(P)$ gravity is being widely used to study late-time acceleration, black holes, and inflation. Recently, tight constraints on $f(P)$ gravity have been reported from the energy conditions \cite{bhattap}. \\
It is well known that there is an asymmetry in the matter content of the Universe where most of the objects we observe are composed of matter while we do not see any such objects composed primarily of antimatter. According to the predictions of the standard big bang cosmological model, the Universe must have produced equal amounts of matter and antimatter only to be annihilated moments later. It is now believed that a minor but significant asymmetry surfaced between matter and antimatter which converted a small fraction of antimatter into the normal matter before they could annihilate. Thus, a tiny residual of matter remained and formed all the structures in the Universe. This phenomenon is also called baryon asymmetry or baryogenesis. Gravitational baryogenesis is one of the most popular theoretical frameworks to explain baryogenesis in great detail \cite{pdu3}. In this framework, by employing the famous Sakharov conditions \cite{pdu4}, a CP-violating interaction was proposed which gives rise to a baryon asymmetry naturally.\\
In this work, we studied gravitational baryogenesis in the framework of $f(P)$ gravity to understand the applicability of this class of modified gravity in addressing the baryon asymmetry of the Universe. For the analysis, we set $f(P) = \alpha P$ where $\alpha$ is the model parameter. We found that in $f(P)$ gravity, the CP-violating interaction acquires a modification through the addition of the nontopological cubic term $P$ in addition to the Ricci scalar $R$ and the mathematical expression of the baryon-to-entropy ratio depends not only on the time derivative of $R$ but also the time derivative of $P$. Additionally, we also investigate the consequences of a more complete and generalized CP-violating interaction proportional to $f(P)$ instead of $P$ in addressing the baryon asymmetry of the Universe. For this type of interaction, we report that the baryon-to-entropy ratio is proportional to $\dot{R}$, $\dot{P}$ and $f^{'}(P)$.  We report that for both of these cases, rational values of $\alpha$ and $\chi$ generate acceptable baryon-to-entropy ratios compatible with observations.

\section*{Acknowledgments}

We thank an anonymous reviewer for the helpful comments.

\end{document}